\documentclass[fleqn,usenatbib]{mnras}
\usepackage{mathptmx}

\usepackage[T1]{fontenc}

\DeclareRobustCommand{\VAN}[3]{#2}
\let\VANthebibliography\thebibliography
\def\thebibliography{\DeclareRobustCommand{\VAN}[3]{##3}\VANthebibliography}

\usepackage{graphicx}	
\usepackage{amsmath}	

\newcommand{\civ}{C~{\sc iv}}

\newcommand{\ovi}{O~{\sc vi}}

\newcommand{\siiv}{Si~{\sc iv}}
\newcommand{\heii}{He~{\sc ii}}

\newcommand{\La}{L$\alpha$}
\newcommand{\mgii}{Mg~{\sc ii}}
\newcommand{\feii}{Fe~{\sc ii}}

\newcommand{\swift}{SWIFT}
\newcommand{\hst}{HST}
\newcommand{\cloudy}{{\it cloudy}}

\newcommand{\MBH}{M$_{\rm BH}$}
\newcommand{\Msun}{M$_{\odot}$}
\newcommand{\cosstis}{{\rm COS-STIS}}
\newcommand{\ergs}{erg~s$^{-1}$}
\newcommand{\Ledd}{L$_{\rm Edd}$}

\newcommand{\tdamp}{$\tau_{\rm damping}$}

\newcommand{\rhalf}{\ifmmode r_{\rm 1/2} \else $r_{\rm 1/2}$\fi}
\newcommand{\Rg}{\ifmmode R_{\rm g} \else $R_{\rm g}$\fi}
\newcommand{\Mrdot}{\ifmmode \dot{M}\left(r \right) \else $\dot{M}\left(r \right)$\fi}
\newcommand{\Medddot}{\ifmmode \dot{M}_{edd} \else $\dot{M}_{edd}$\fi}
\newcommand{\Moutdot}{\ifmmode \dot{M}_{\rm out} \else $\dot{M}_{\rm out}$\fi}
\newcommand{\Mindot}{\ifmmode \dot{M}_{\rm in} \else $\dot{M}_{\rm in}$\fi}
\newcommand{\Mwinddot}{\ifmmode \dot{M}_{\rm wind} \else $\dot{M}_{\rm wind}$\fi}

\title[Disk winds]{Disk winds spectral energy distribution and the variability timescale of active galactic nuclei}

\author[Hagai Netzer]{
Hagai Netzer,$^1$\thanks{E-mail: hagainetzer@gmail.com}
\\
$^{1}$ School of Physics and Astronomy, Tel Aviv University, Tel Aviv 69978, Israel 
}

\date{Accepted XXX. Received YYY; in original form ZZZ}

\begin{document}
\label{firstpage}
\pagerange{\pageref{firstpage}--\pageref{lastpage}}
\maketitle


\begin{abstract}
        Accretion disk winds are commonly observed in active galactic nuclei (AGN). The winds may be associated with the changing spectral properties of such sources, yet such connections have not been studied in detail so far.
        This paper presents detailed calculations of accretion disk winds and their impact on the observed spectrum of AGN, emphasizing recent observations of Mrk\,817. The model consists of a radial and time-dependent mass outflow rate
        with a half ejection radius of 50\Rg\ and a variability timescale of 100 days.
        The resulting spectral energy distribution (SED) is characterized by a large drop in the ionizing luminosity and less significant changes in the optical luminosity. The time-dependent intensities of the broad emission lines and the spectrum emitted by the wind material reflect these variations.
        For Mrk\,817, the variability timescale of the 1500-5500\AA\ continuum, the thermal time of the disk at the locations where most of this radiation is emitted, and the range of radii over which most of the mass outflow is taking place are all in agreement with the black hole mass and the radial-dependent accretion rate through the disk. This suggests a general connection between time-dependent disk winds, SED shape, thermal timescale, and optical-UV variability in AGN.

\end{abstract}


\begin{keywords}
accretion, accretion discs –galaxies: active –galaxies: nuclei –quasars: general
\end{keywords}



\section{Introduction}

Active galactic nuclei (AGN) are characterized by large-amplitude, wavelength-dependent variations that are more easily observed in low-luminosity sources \citep[see][and references therein]{Kelly2009,MacLeod2010,Arevalo2024}. The variability affects the optical-UV continuum, the broad emission lines, and the X-ray radiation from the source. Some of the best examples of this type were obtained in recent years by combining ground-based and space-borne observations of low-redshift AGN. The AGN STORM 1 collaboration followed NGC\,5548 \citep[see][]{deRosa2015}, and the AGN STORM 2 collaboration  monitored   Mrk\,817 \citep[][]{Kara2021}. 

Both AGN STORM collaborations found indications for powerful mass outflow from the vicinity of the central black hole (BH) \citep[e.g.][]{Zaidouni2024}, likely connected to the central accretion disk. In Mrk\,817, there are also indications for a changing spectral energy distribution (SED) and a bluer continuum during the higher luminosity phases  
\citep[see][hereafter N24]{Netzer2024}.

In this paper, I present a comprehensive analysis of disk winds and their likely connections to SED variations, optical-UV variations, and the relationship between the variability timescale and the thermal time of the inner accretion disk. Most of the detailed examples apply to the case of Mrk\,817, due to the superb data provided by the AGN STORM 2 campaign. Still, there are general implications to disk winds in other individual sources, a possible connection to mass outflow in large AGN samples \citep[e.g,][]{Matthews2023,Temple2021}, and microlensing observations of high-redshift, high-luminosity AGN.

 Section\,2 of the paper summarizes the evidence for disk winds in AGN and presents a generic wind model designed to investigate the spectral variations resulting from such winds.  In section\,3, I present radiation pressure-confined cloud models used to study the emission line variations in the broad line region (BLR), and the expected spectral signature of the wind. Section\,4 discusses the implications of disk winds to wavelength-dependent spectral variations in Mrk\,817, and the connections between variability timescales, the thermal time of the accretion disk,  and other disk properties. Section\,5 presents the conclusions of the paper. 

Most of the observations discussed in this paper were presented in several AGN STORM 2 papers \citep[e.g.][]{Kara2021,Cackett2023,Homayouni2023,Partington2023,Homayouni2024,Zaidouni2024}. Quoted luminosities are based on z=0.0315 and
the following cosmological parameters: $H_{0}=$70~km~s$^{-1}$~Mpc$^{-1}$,
 $\Omega_{\Lambda}=0.7$, and $\Omega_{M}=0.3$.

\section{Disk winds and SED variations in AGN}
\label{SED_variations}
\subsection{SED variations in large AGN sample}

All type-I AGN show flux variations at most observed wavelengths. 
This topic has been extensively discussed in the astronomical literature with various attempts to correlate the measured timescales to disk and radio source properties \citep[see][and references therein]{Kelly2009}.
\cite{MacLeod2010} analyzed $\sim$9000 quasars in SDSS Stripe 82 and modeled their luminosity variations as a damped random walk (DRW) or CAR(1) process. They found that the asymptotic
long-term rms variability increases with decreasing luminosity. The rms also increases with BH mass without correlating with the decreasing luminosity. The characteristic variability timescale increases with wavelength and BH mass, and the variability amplitude is anti-correlated with the Eddington ratio. There are additional claimed correlations that are harder to confirm. 

Correlations of SED shape with optical-UV variations require wide-band observations and are harder to quantify. Well-confirmed cases of individual sources, like NGC\,5548 \citep[see, e.g.,][]{Wamsteker1990,Maoz1993,Korista1995}, as well as studies of large AGN samples \citep[see][]{Vandenberk2004,Sun2014}, show that in many cases the object becomes ``bluer when brighter''. 
According to \cite{Sun2014}, such color changes are most noticeable on short-time scales, which was interpreted as indicative of thermal fluctuations in the inner, hotter part of the central accretion disk.

\subsection{SED variations in Mrk 817}

The recent study of Mkn\,817 by the AGN\,STORM\,2 collaboration \citep{Kara2021} provides a superb benchmark to test SED variations because of the wide wavelength band provided by \hst, \swift, and several ground-based observatories, and due to the long duration of the campaign (about 500 days). Paper\,X in the AGN\,STORM\,2 series (N24) demonstrated changes in the shape of the SED of the source by comparing two \hst/\cosstis\ spectra that cover the 1100-9000\AA\ wavelength range and are separated by about 100 days. The 1100\AA\ continuum bands of the two spectra differ in flux by about a factor of two, while the difference between the galaxy-corrected fluxes at 9000\AA\ is only about 20\%. 
 Here, I utilize these observations to further investigate the SED variations of the source. This was done by subtracting the host galaxy contribution from the \swift\ observations 
assuming zero contribution to the U, B, and V bands in the \cosstis\ spectra. 

\begin{figure} \centering      
\includegraphics[width=0.95\linewidth]{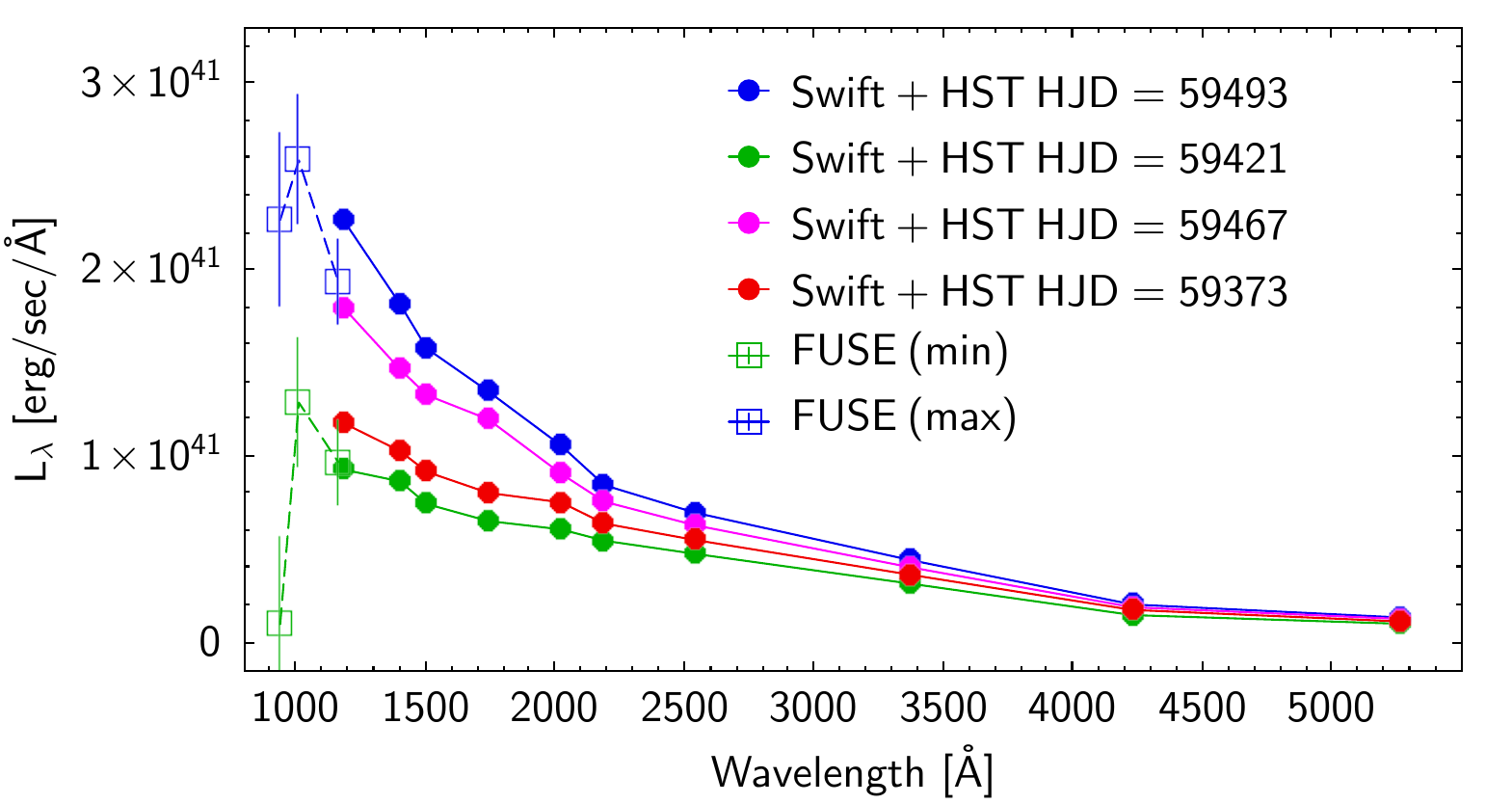}
        \caption{Host-galaxy subtracted \hst-\swift\ SEDs covering the
        entire luminosity range observed during the AGN\,STOR\,2 campaign. I also show the brightest and faintest FUSE observations obtained on 18.2.2001 and 23.12.2000, respectively. 
        }
        \label{hst_swift_spectra}
\end{figure}

Fig.~\ref{hst_swift_spectra} shows four host-galaxy subtracted combined \hst-\swift\ spectra that cover the entire luminosity range during the campaign. The SEDs obtained by this procedure are bluer during the higher luminosity epochs. I also show mean fluxes obtained by FUSE observations during 2000 and 2001 (see 
\cite{Scott2004}, and information in the figure caption and in https://archive.stsci.edu). 
 The source luminosity during the FUSE observations was similar to that observed during the AGN STORM 2 campaign. The highest flux FUSE spectrum shows a strong \ovi\ emission line, and the lowest flux spectrum shows an absorption feature at the redshifted Lyman limit, possibly indicating an optically thick line-of-sight wind.

\subsection{Time-dependent mass outflow from AGN disks}
\subsubsection{Disk wind models}
As explained earlier, powerful winds from the vicinity of the central BH are associated with AGN activity, with characteristic signatures in the X-ray and the blue wings of several broad absorption lines like \civ\ and \siiv. While the origin of such winds is still debated, there are indications that the distances of the absorbing material are of order 100-1000 gravitational radii (\Rg).

Mass outflow affects the mass accretion rate and, hence, the luminosity of the accretion disk. Such outflows can originate from the outer parts of the disk, where their influence on the observed flux is on very long, viscous time scales. They can also be ejected much closer to the BH, where much shorter time scales are involved, with significant implications for the shape of the SED.  
\cite{Slone2012} (hereafter SN12) considered three types of disk winds with various accretion rate normalizations. They showed that winds with a characteristic half-radius of 50-100 \Rg\ can cause a significant steepening of the observed UV SED, compared with the SED of a standard thin accretion disk \citep[e.g.,][hereafter SS73]{Shakura1973}. This may be related to the steep short-wavelength part of the spectrum of many AGN. 
\cite{Laor2014} (hereafter LD14) considered a more specific wind that originates from the parts of the disk where the conditions are similar to 
those observed in O-type stars with powerful stellar winds. For a BH mass of order $ 10^8$\Msun, and an accretion rate of about 0.1\Msun/yr, this corresponds to about 50Rg.
The effect of such winds on the observed SED is similar to several cases studied by SN12. Section 4.3 provides more details on the LD14 model.

This paper's primary goal is to investigate the possibility that powerful disk winds are the source of short-timescale SED variations reported in AGN. The SN12 and LD14 wind calculations are not intended to explain such variations in slowly evolving accretion disks. Nevertheless, the lack of suitable numerical calculations motivates the investigation of their predictions. 

To calculate the modified SED, I follow the procedure described in SN12 and assume that the integrated mass accretion rate at radius $r$ is  given by
\begin{equation}
 \Mrdot=\Mindot+\int_{r_{in}}^{r}d\Mwinddot(r),
\label{eq:Mdot_Mwinddot}
\end{equation}
and reaches its maximum at $r_{out}$. An additional parameter, $\rhalf$, is the radius at which $50\%$ of the total mass in the wind has been ejected.
SN12 experimented with three types of winds: 
\begin{enumerate}
 \item   
A self-similar wind,
\begin{equation}
 \frac{d\Mwinddot(r)}{dr}=\frac{a_1}{r},
\label{eq:wind1}
\end{equation}
where $a_1$ is a normalization parameter and depends on $\Mindot$ and $\Moutdot$.
\item 
A step function wind,
\begin{equation}
 \frac{d\Mwinddot(r)}{dr}=b_1\left(1-\left[1+\exp\left(-2b_2\left(r-r_m\right)\right)\right]^{-1}\right),
\label{eq:wind2}
\end{equation}
where $b_1$, $b_2$  and $r_m$ are parameters which determine $\rhalf$.
\item
A linear wind,
\begin{equation}
 \frac{d\Mwinddot(r)}{dr}=c_1-c_2\cdot r,
\label{eq:wind3}
\end{equation}
where $c_1$ and $c_2$ are chosen such that $\rhalf$ has its chosen value.
\end{enumerate}
According to SN12, for a chosen $\rhalf$, type-ii and type-iii winds produce similar SEDs, while type-i winds can be significantly different. Graphical illustrations of all three cases are shown in SN12, Figure\,2. Most of the results shown below assume
$\rhalf$=50\Rg\ and a type-ii wind.

The calculated disk SEDs shown in this paper use the SN12 code. They include GR corrections and a modified 
balance between the gravitational potential and torque energies within the disk.
They also take into account Comptonization in the disk atmosphere. In the following, I refer to two distinct cases, ``wind'' and ``no-wind''. The first is associated with the lowest luminosity observed in Mrk\,817, and the second with the highest luminosity phase. However, there is no clear distinction between this situation and the alternative case of ``strong wind'' and ``weak wind''. Moreover, the nature of the instabilities leading to luminosity changes is unknown. A possible scenario may involve a disk with no wind and occasional onset of a mass outflow event, leading to a lower luminosity phase. Alternatively, a steady mass outflow episode can occasionally transition to a lower mass outflow rate, leading to a more luminous phase.   

\subsubsection{Comparison with SED observations}

\begin{figure} \centering      
\includegraphics[width=0.95\linewidth] 
 {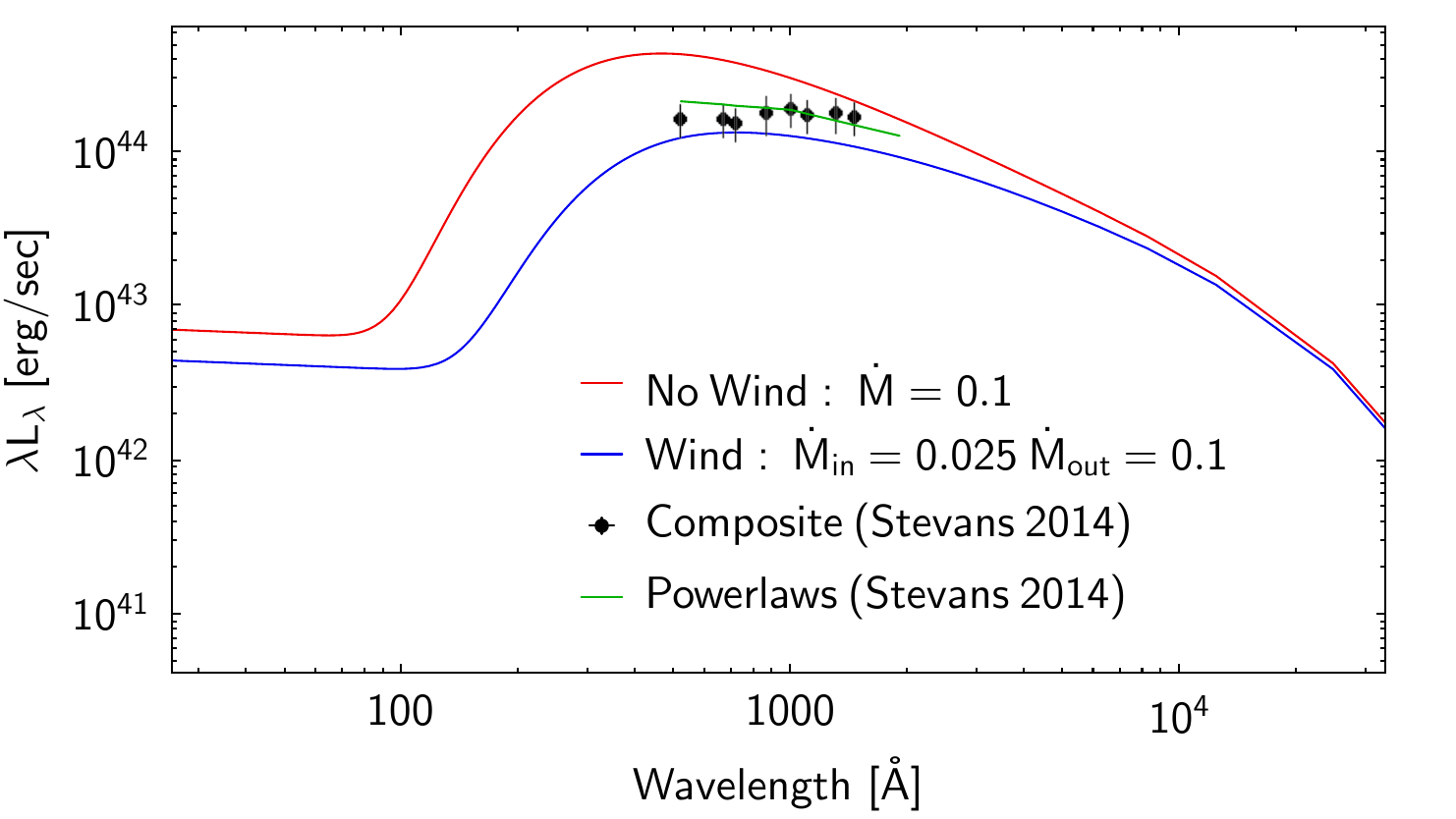}
        \caption{
        Calculated SEDs for the brightest and faintest spectra of Mrk\,817 during the AGN STORM 2 campaign. 
      $\dot{M}$ is in units of \Msun/yr.
	 Points with error bars represent the normalized composite spectrum of
	 Stevans (2014).
                The green broken line shows a combination of two power laws
                representing a third of the sources with the steepest rising
               spectrum. These are primarily low-redshift AGN similar to 
                Mrk\,817.
 }
        \label{m817_disk_xray_SEDs}
\end{figure}

Using
\MBH=$10^{7.6}$\Msun, \Moutdot=0.1\Msun/yr, \Mindot=0.1\Msun/yr,
$a=0.7$ and $\cos i=0.75$, I calculated various wind and no-wind SEDs designed to fit the observations of
Mrk\,817. The highest luminosity SED is the model used in N24 to fit the \cosstis\ spectrum obtained on HJD=2459322.
In this case, the luminosity and SED shape are very similar to those at HJD=2459493. The SED of the disk was combined
with an X-ray source similar in shape to the one discussed in N24, with
$\alpha_{OX}=-1.4$.
The model is shown with a red line in Fig.~\ref{m817_disk_xray_SEDs}. 
The assumption is that during this epoch, the effect of the wind is smaller than in all other cases. The lowest luminosity case considered here was obtained by keeping the same \Moutdot, assuming a type-ii wind and \rhalf=50\Rg,  and reducing \Mindot\ by a factor of 4. This SED is shown with a blue line in Fig.~\ref{m817_disk_xray_SEDs}. The 1000-9000\AA\ luminosity is similar to the one observed at HJD=2459545. 

 Fig.~\ref{m817_disk_xray_SEDs} also shows some of the results of \cite{Stevans2014}, who analyzed HST/COS observations of 159 AGN with different redshift and luminosity and used them to illustrate typical AGN continuum spectra. The dots with error bars show the normalized composite SED from that paper. The green, broken power-law represents low redshift AGN with
 $\lambda L_{\lambda}$(1100\AA)$ \approx 10^{44.3}$\ergs, similar to Mrk\,817. The SED of such sources is between the high and low luminosity SEDs considered in this paper.   

\begin{figure} \centering      
\includegraphics[width=0.95\linewidth] 
 {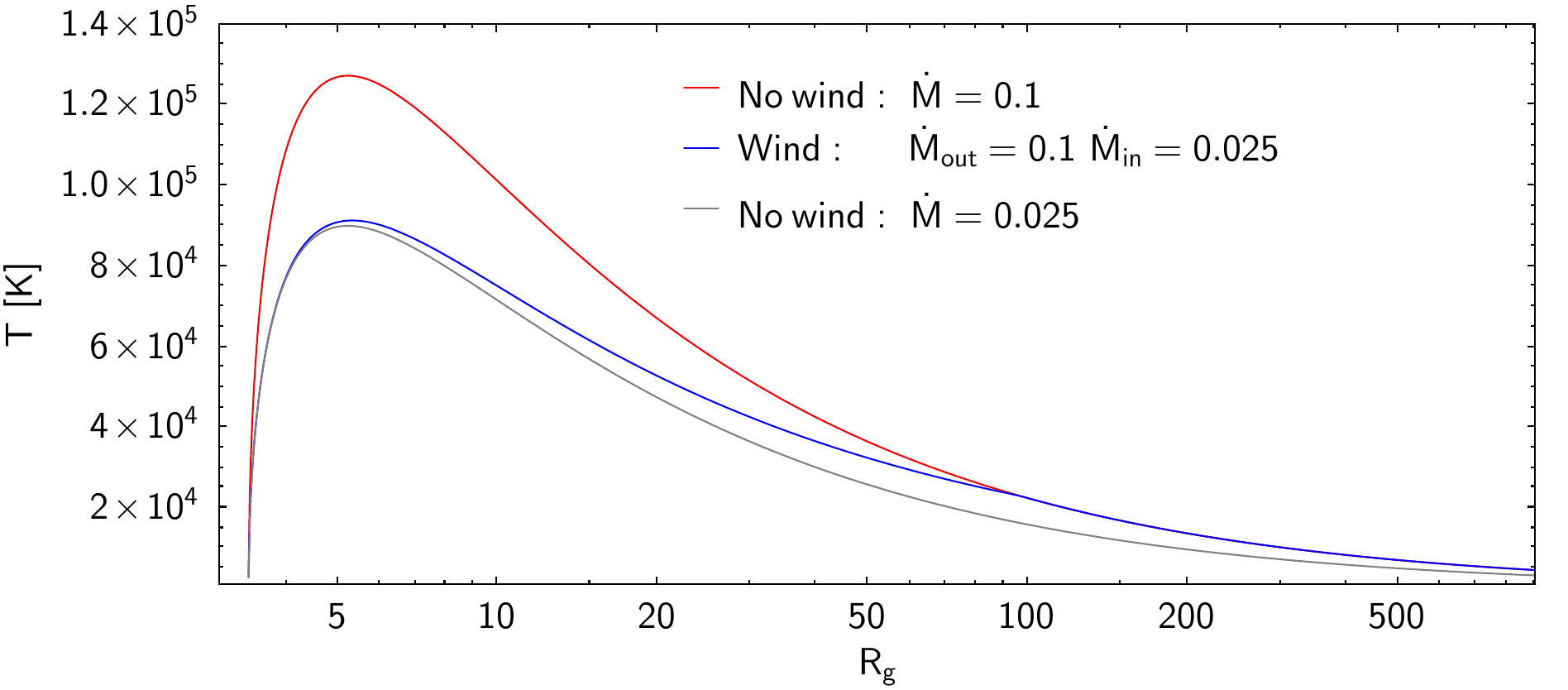}
        \caption{Calculated disk temperatures for the two SEDs shown in Fig.~\ref{m817_disk_xray_SEDs}, using the same color scheme, and a third case, drawn in a grey line, where $\dot{M}_{\rm in}=\dot{M}_{\rm out}=0.025$~\Msun/yr.
        }
        \label{disk_temperatures}
\end{figure}

To illustrate the changes in disk properties, I show in Fig.~\ref{disk_temperatures} the run of the disk surface temperature in the high and low-luminosity cases. As expected, including a strong disk wind results in a significant temperature drop in the inner parts of the disk. The drop is demonstrated by the third plotted curve, which shows the temperature profile for a case where 
$\dot{M}_{\rm in}=\dot{M}_{\rm out}=0.025$~\Msun/yr. I also examined the dependence on \rhalf\ and found slight differences between \rhalf=25\Rg, \rhalf=50\Rg, and \rhalf=100\Rg. In the remainder of the paper, I focus on the case of \rhalf=50\Rg.

The radius-dependent accretion rate has a significant impact on the total disk luminosity and a more pronounced effect on its ionizing luminosity.
For example, keeping 
\Moutdot\ as is and reducing \Mindot\ by a factor of four results in a reduction of the total disk luminosity by approximately 63\% and a reduction of roughly 76\% in ionizing luminosity.

\subsubsection{What is the mass outflow rate in Mrk~817?}

The mass outflow rate required to fit the faintest observed SED during the AGN\,STORM\,2 campaign is 0.075\Msun/yr. Do the observations support this?

The mass outflow rates derived from the observations of the broad absorption lines of \civ\ and \siiv\ cannot provide an answer to this question because the covering factor is unknown, the optical depths of the lines change over time, and there are periods of very weak absorption as well as periods of saturated absorption \citep[see][]{Partington2023}. 
Analyzing the high and low-resolution X-ray observations reported in \cite{Partington2023} and \cite{Zaidouni2024} can provide more realistic estimates.

\cite{Zaidouni2024} carried out a detailed analysis of several episodes where the source was bright enough to use the high-resolution RGS instrument on board XMM-Newton. The analysis revealed a multi-component absorber with strong and broad X-ray lines whose profiles are very similar to those of the \civ\ line observed by \hst. It also provided reliable column density and distance measurements, allowing an estimate of the mass outflow rate, assuming mass-conserving flow. Under these assumptions,
\begin{center}    
$\dot{M}/C_f > 4 \pi R(wind) m_p N_H v(flow) \approx 0.06 $\,\Msun/yr, 
\end{center}
where $C_f$ is the covering factor of the flow, 
$R(wind)=5.7 \times 10^{15}$\,cm, $m_p$ is the mass of the proton, 
$v(flow)=5200\, {\rm km s^{-1}}$, and the absorber column density is $N_H=6.6 \times 10^{22} {\rm cm}^{-2}$.

The disk wind model used in this paper (section~\ref{disk_winds}) assumes several mass ejection episodes separated by approximately 100 days, observed as X-ray absorption episodes at $R(wind)=10^{16}$\, cm. I utilize the data and models provided in \cite{Partington2023} to analyze the time-dependent column density variations.
The outflow velocity incorporates measurements from \cite{Zaidouni2024} and adjusts for the angle to the line of sight, assuming the wind is emitted perpendicular to the disk's surface.  
Such winds could reach our line of sight in about 100 days if ejected from the disk's surface at 1000\Rg. A similar time scale is obtained by assuming that the wind is ejected at the local escape velocity some  100\Rg\ from the BH and moves in a radial direction without additional acceleration, where for most of the time, it does not block our line of sight to the X-ray source and the BLR line of sight to the ionizing radiation emitting parts of the disk.  
Both scenarios are compatible with the assumption that the wind is observed when it reaches a distance of approximately $10^{16}$\ cm. 

The mean X-ray column density measured by \cite{Partington2023} is $1.1 \times 10^{23}$\,cm$^{-2}$. About half comes from periods when the column density is $5\times 10^{22}$\,cm$^{-2}$,  and the other half from periods when the column density is
$1.5 \times 10^{23}$\,cm$^{-2}$. There are no periods with very low column density, suggesting a covering factor of the X-ray corona larger than 0.5. Three high-column density episodes each year, lasting approximately 60 days each. The high outflow velocity indicates that each high column density episode is observed only once.  Given these estimates, the mass outflow rate during the AGN STORM 2 campaign is about 0.1 \Msun/yr.  This is consistent, within the uncertainties, with the accretion rate at the outer parts of the disk. It is also consistent, within the uncertainties on the velocity, distance, and column density, with the \cite{Zaidouni2024} estimate of $\dot{M}$ if $C_f \approx 0.25$. 

\section{Disk winds and time-dependent BLR spectra}
\label{disk_winds}
\subsection{Modeling the maximum and minimum mass outflow rate episodes in Mrk~817}
\subsubsection{Wind induced spectral variations}

The broad emission and absorption lines in the spectrum of Mrk\,817 have been discussed in several AGN STORM 2 collaboration papers. All the broad emission lines change their flux on time scales of 3-30 days, following the optical-UV continuum variations. Significant diffuse continuum (DC) variations are also observed on shorter time scales. 
The diffuse line and DC variations also affect the measured interband continuum lags. Detailed measurements of such variations are presented in \cite{Cackett2023}, and extensive modeling is presented in N24.

N24 presented several flared-shaped BLR models for Mkr\,817. The models assume radiation pressure-confined (RPC) clouds extending from the central BH over 2-122 light days. They successfully explain the intensity and lags of most emission lines as well as the observed interband continuum variations. The paper addressed the observed winds and argued that they are necessary to explain several episodes of time-dependent emission that have been observed.

The N24 models considered only two SEDs resulting from two constant accretion rate disks: $\dot{M}=0.1$~\Msun/yr, and  $\dot{M}=0.05$~\Msun/yr. The shapes of the two SEDs are very similar and do not differ much from a case where the second SED is simply the first SED multiplied by 0.5.
As shown in Fig.~\ref{m817_disk_xray_SEDs}, such SEDs differ significantly from those considered here. Below, I present the BLR spectra predicted by the disk wind scenario.

The two cases considered here represent wind and no-wind BLR models. The no-wind SED (hereafter BLR1, L$_{\rm max}$) is identical to the case defined as "model~3" in N24. The windy disk model (BLR2, L$_{\rm min})$ is ionized by the suppressed SED presented earlier, with 
$\dot{M}_{\rm in} =0.025 \dot{M}_{\rm out}$. The inner boundary of the two BLRs is at $10^{16.25}$ cm. The outer boundaries are at $10^{17.50}$ and $10^{17.3}$ cm, respectively, reflecting the (somewhat simplistic) assumption that the BLR extends to the sublimation radius of graphite grains. 
Turbulence velocity of 30 km/sec, and a
  covering factor given by $dC(r) \propto r^{-\beta}$ where $\beta=2$ and $C(r=10^{17.5}\,cm)=0.2$, are assumed.

The L$_{\rm min}$ phase includes emission by the outflowing gas. This is assumed to extend over the distances of $10^{15.25-15.50}$\,cm, in a direction different from the BLR gas, and a total covering factor and column density identical to those of the BLR 
\footnote{In section 4.1 I use a more extended wind covering all distances up to 10$^{16}$\,cm. The emitted spectra of the two winds are very similar.}

The BLR gas ionization and electron temperature are calculated with the photoionization code \cloudy\ \citep{Ferland2017}. There are three changes compared with the N24 calculations. First, the gas metallicity is three times solar, compared with the solar metallicity gas assumed in N24. This was introduced to check the dependence on gas composition and was found to be of minor importance. Second, the present paper uses version 23.01 of \cloudy\ compared with version 17.02 in N24. The differences between the two are very small for the calculations presented here. Third, and more important, is the use of the full \feii\ ion model provided in \cloudy\ compared with the simplified model used in N24. This results in enhanced emission over the wavelength range of 2000-3000\AA, and provides a better fit to the observed spectra of Mrk\,817.

The calculated intensities of most broad emission lines during the high luminosity, no-wind phase agree with observations. The exceptions are the Balmer lines and the \mgii$\lambda 2798$\AA\ line. This well-known deviation is unexplained and probably results from the simplified escape probability approach used in calculating the line transfer.

\begin{figure} \centering     
\includegraphics[width=0.95\linewidth] 
{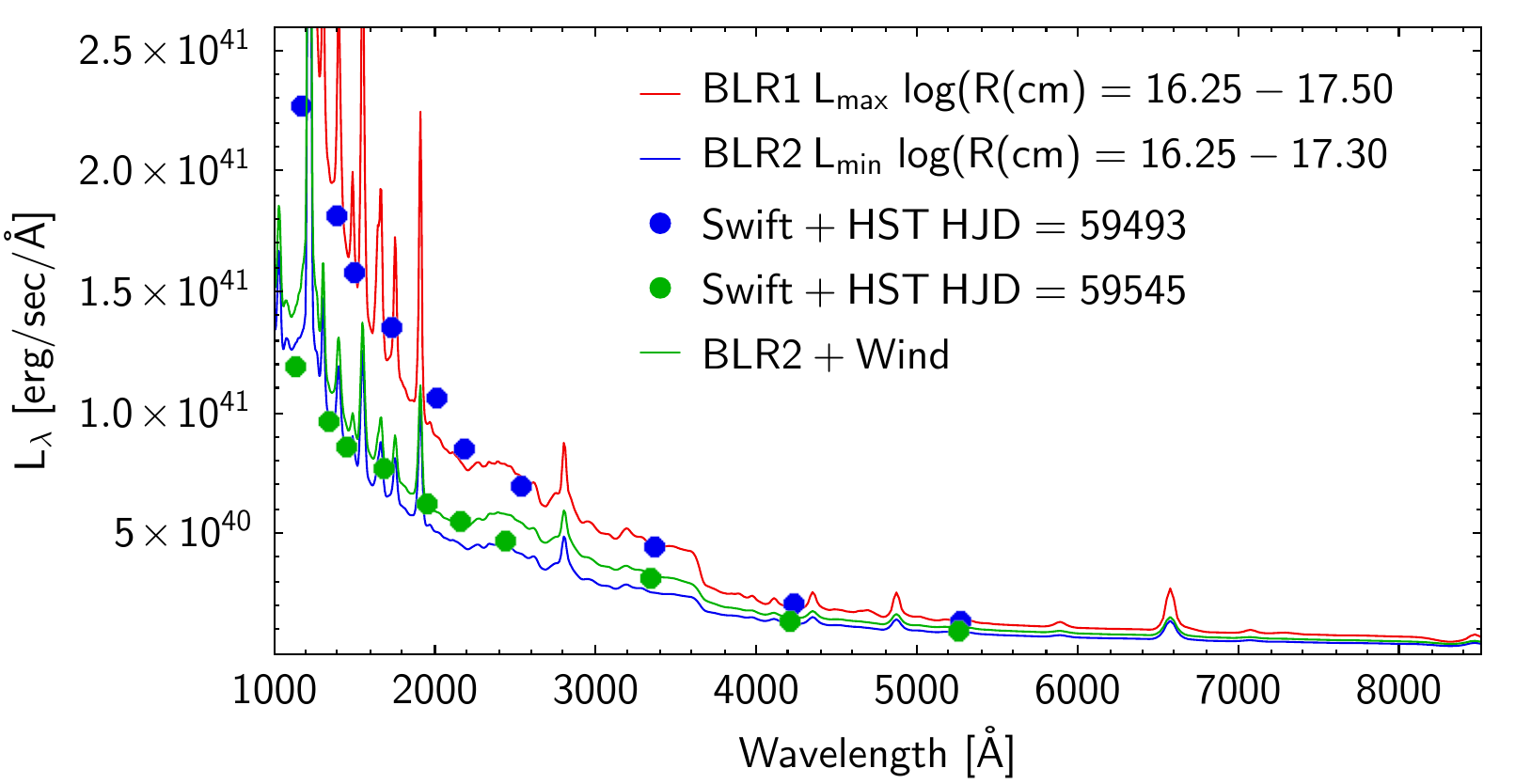}
 \caption{ A comparison between the highest and lowest luminosity \hst/\swift\ observations and the wind BLR models discussed in this paper.
 }
 \label{model_vs_observations}
\end{figure}

Fig.~\ref{model_vs_observations} shows two calculated models for the highest and lowest luminosity epochs, adding the incident disk continuum with the diffuse gas emission, and compares them with the  \hst+\swift\ spectra. The comparison does not extend beyond approximately 8000\AA\ because of possible contamination of the longer wavelength spectrum by dust emission from the torus (see N24). 
The high luminosity no-wind model agrees well with the observations. The low luminosity windy model includes both emission from the BLR and the outflowing gas. Their combination agrees with the observed \hst+\swift\ spectrum during this phase. 

\begin{figure} \centering      
\includegraphics[width=0.95\linewidth]
{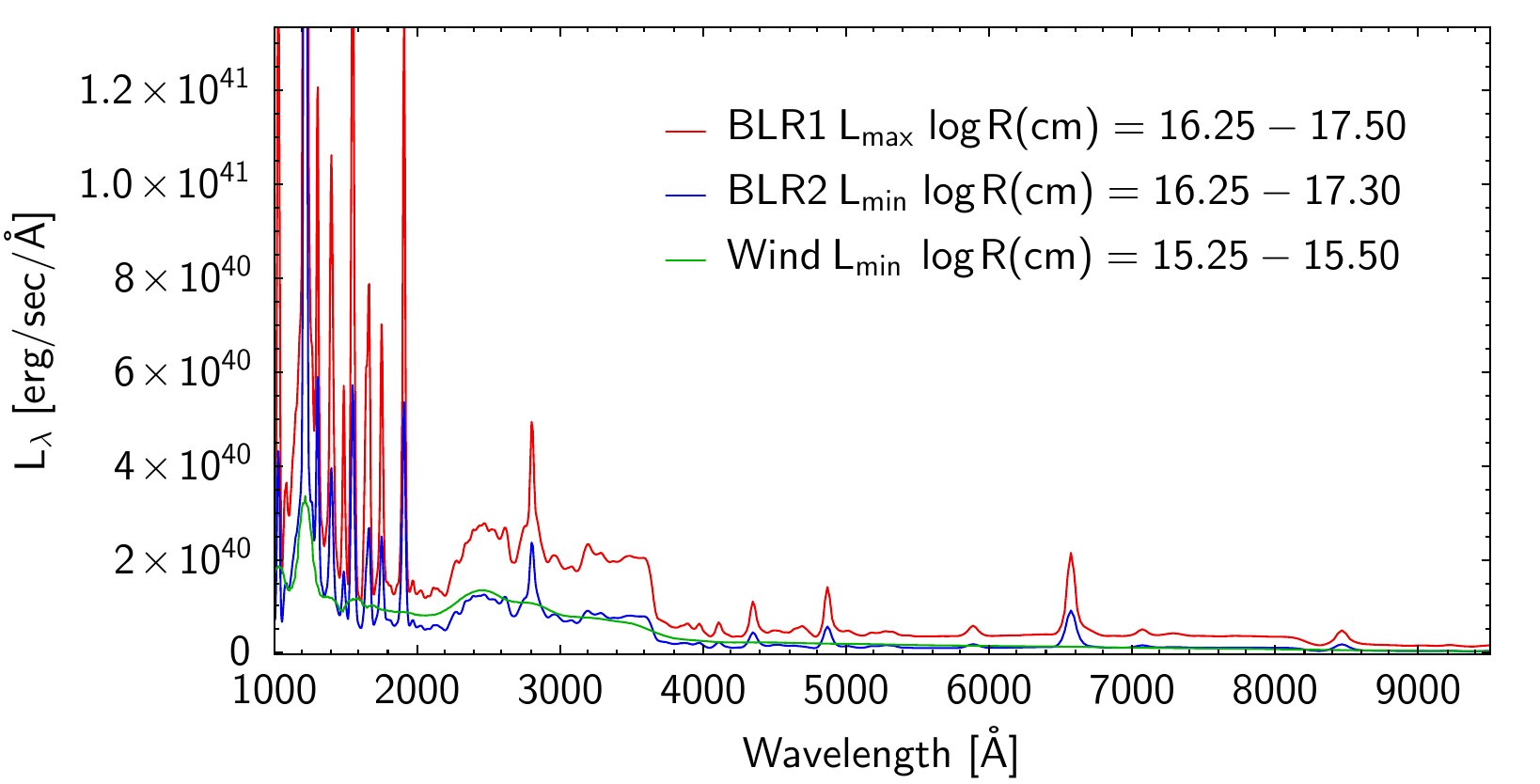}
        \caption{Calculated diffuse emission of two BLRs, and a disk wind, exposed to the wind and no-wind SEDs discussed in this paper.  
        }
        
        \label{wind_no_wind_BLRs}
\end{figure}

\begin{figure} \centering      
\includegraphics[width=0.95\linewidth]
{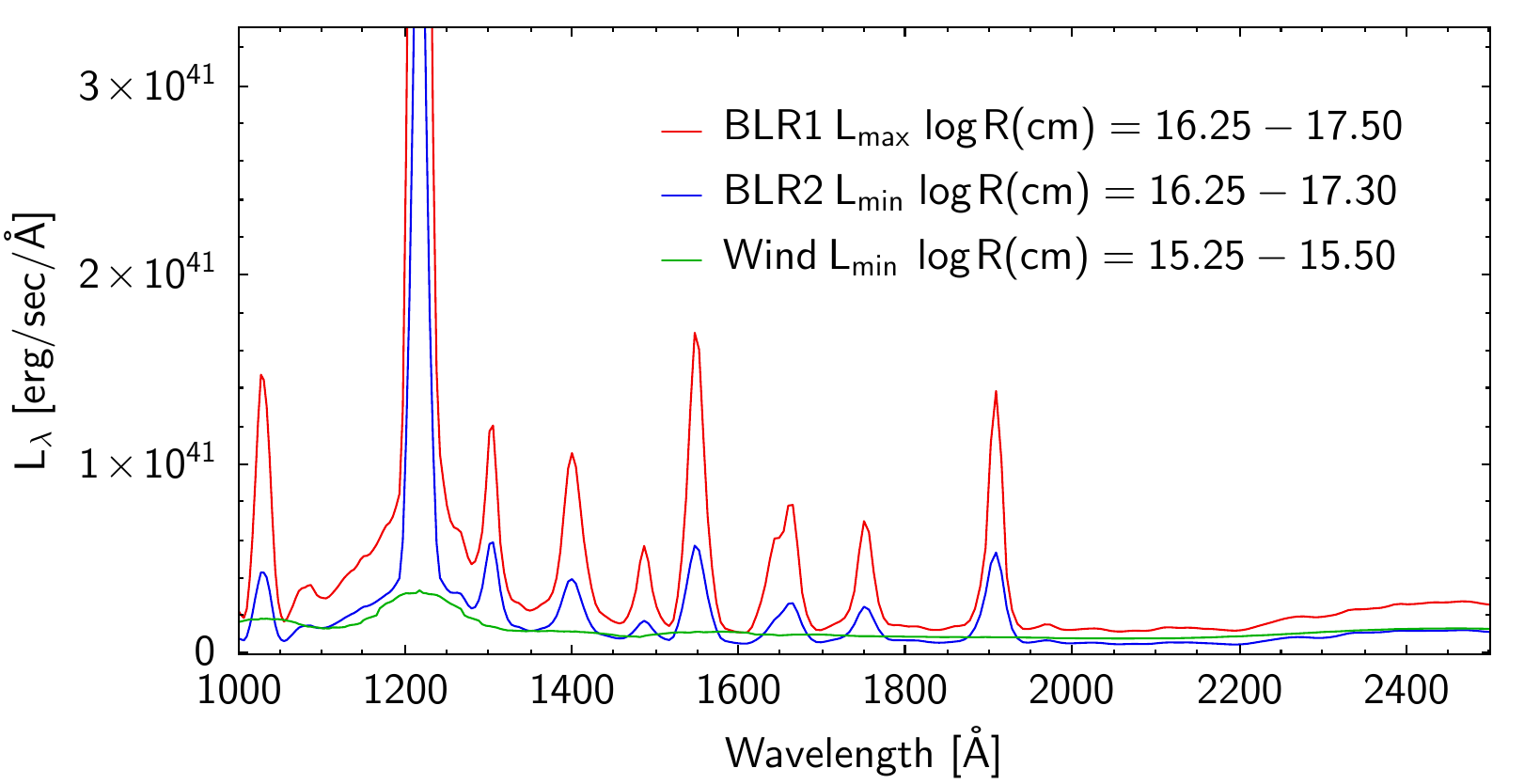}
        \caption{An expanded view of Fig.~\ref{wind_no_wind_BLRs} showing the short wavelength part.   
        }
        
        \label{wind_no_wind_BLRs_expanded}
\end{figure}

 Fig.~\ref{wind_no_wind_BLRs} shows the calculated diffuse emission of the two BLR models, and the gas in the wind, on an expanded scale. 
 The most noticeable differences between the two BLR spectra are the large drops in line intensity and mean ionization level. For example, the \heii$\lambda 1640$\AA\ line luminosity drops by about a factor of four, and the decrease in DC luminosity is about three. Fig.~\ref{wind_no_wind_BLRs_expanded} shows an expanded view of the short-wavelength part of the same diagram. It illustrates that the kind of wind studied in this paper results in extremely weak broad emission line wings even for the strongest lines such as \civ. This differs from the cases studied by \cite{Matthews2023} and \cite{Temple2021}, which are more relevant to outflowing BLR gas further away from the central BH.

 \subsubsection{Variations in the mean emissivity radii}

\begin{figure} \centering      
\includegraphics[width=0.9\linewidth]
{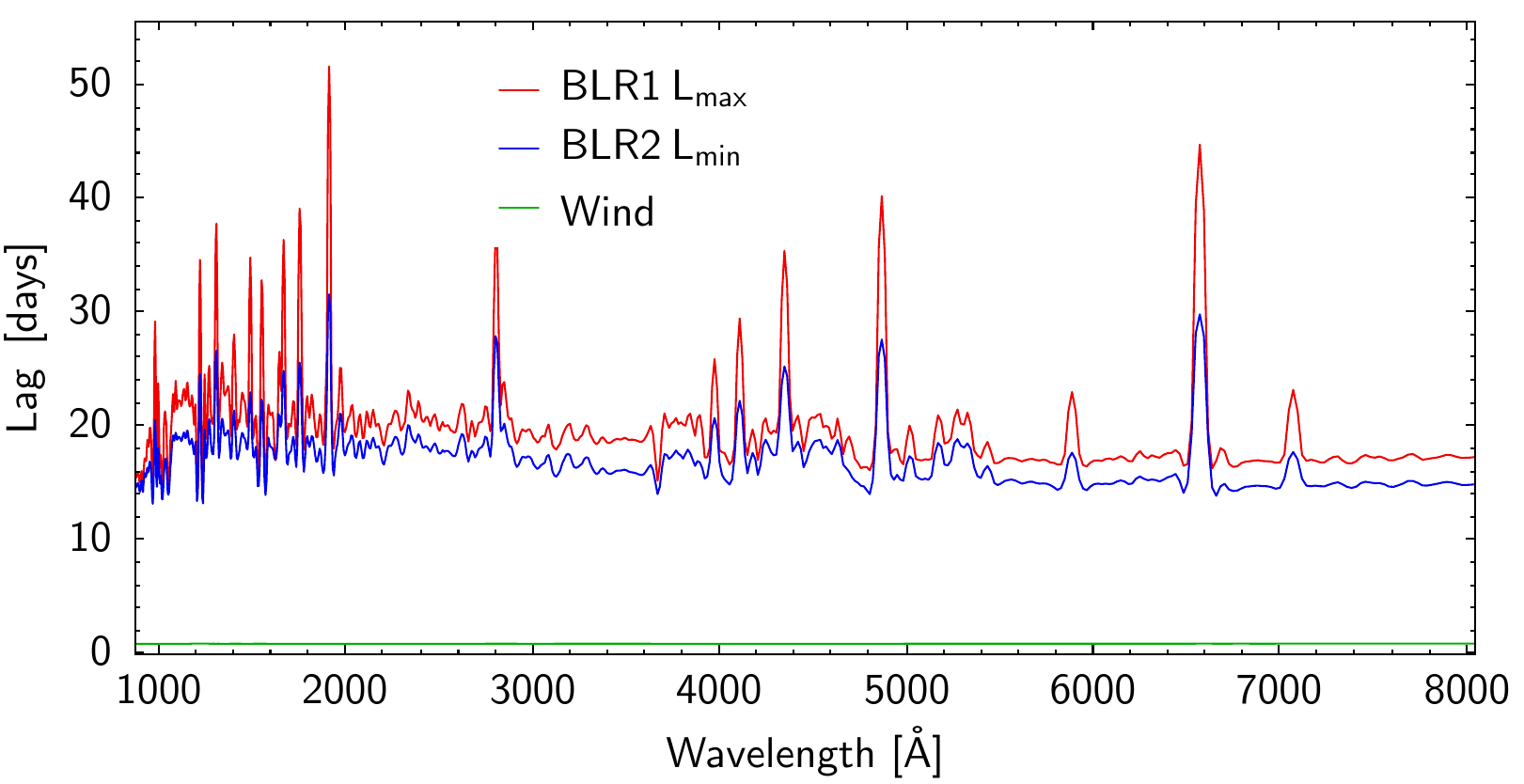}
        \caption{Calculated mean emissivity radii for the wind and no-wind BLR models shown in Fig.~\ref{wind_no_wind_BLRs}. 
        }
        \label{wind_no_wind_MERs}
\end{figure}

 Figure \ref{wind_no_wind_MERs} shows the mean emissivity radii (MER) differences between the wind and no-wind models. Some of these variations arise from the assumed change in \( r_{\text{out}} \), particularly in lines where much of the emission comes from the outer regions of the BLR. Other changes result from a decrease in the level of ionization of the gas.
 For DC emission, the significant drop in luminosity has a minimal impact on the mean emissivity radius, as the spatial distribution of the DC-emitting gas remains quite similar in both cases.  

 \subsubsection{Variations in interband lags}
\begin{figure} \centering      
\includegraphics[width=0.95\linewidth]
{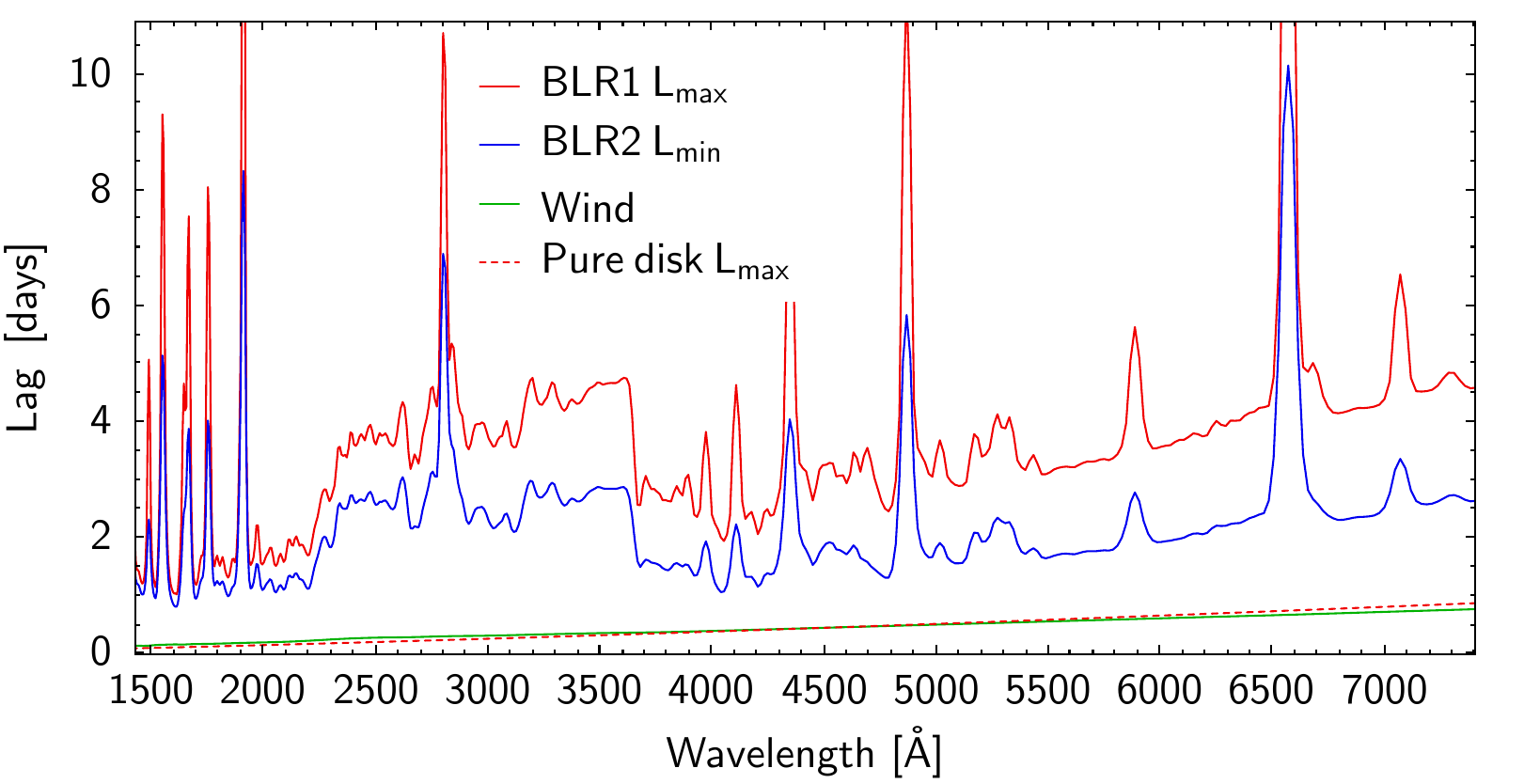}
        \caption{Predicted interband lags due to the onset of a type-ii disk wind with the properties discussed in the paper. 
        }
        
        \label{wind_no_wind_continuum}
\end{figure}

 The predicted interband lags for the wind and non-wind cases are presented in Fig.~\ref{wind_no_wind_continuum}. This topic was thoroughly discussed in N24, where it was demonstrated that "Model~3", which is also the no-wind model used in the present paper, provides a good fit to the lags measured for the entire AGN~STORM~2 campaign (see Fig. 10 in that paper). 
 
 The calculations presented here suggest that the interband lags observed in Mrk\,817 represent two distinct luminosity states of the source. The high luminosity phase (L$_{max}$ in the figure) fits the lags measured during these times. In particular, they agree with "Epoch-1" and "Epoch-3" discussed in \cite{Lewin2024}.
 The onset of the disk wind reduces the ionizing luminosity by four
 and leads to the low luminosity phase, L$_{min}$, consistent with ``Epoch-2'' of \cite{Lewin2024}. The result is a smaller 
 $L_{\rm diffuse}/L_{\rm disk}$ ratio and significantly shorter interband lags. In this scenario, the wind does not shield the BLR, which still emits strong emission lines; instead, it is the main cause of the shortening of the interband lags.
 Separating them through CC analysis is challenging because low and high-luminosity episodes are short (see N24).

\section{Discussion: Disk winds, disk size, and variability timescale}

The results presented in the previous sections focus on the variable observed spectra of Mrk\,817. They show how the onset of a disk wind at distances of less than about 100\Rg, can explain the observed changes in SED shape, in line and continuum lags, and in the time-variable X-ray absorption. Here, I discuss two additional consequences of such winds related to the total observed size of the system concerning microlensing observations and the variability time scales of AGN disks. 

\subsection{Disk winds and microlensing}

Earlier investigations of interband lags in nearby AGN suggested that disk illumination by a variable central X-ray source is the source of the observed continuum lags. This significantly increased the deduced accretion disk size compared to a standard SS73 disk. More recent works, including \cite{Netzer2022} and N24, showed that much of the measure continuum lag is due to line and continuum variations in the BLR. This removed the need to assume extremely large AGN disks.  
Measurements of microlensing events in several high-luminosity, high-redshift AGN provide an independent method for estimating the accretion disk size. The sizes measured at rest wavelengths of about 2500\AA\ were found to be 3-5 times larger than the mean surface brightness (SB) sizes of standard SS73 disks
\citep[see][and references therein]{Motta2017,Morgan2018,Cornachione2020}, and implied a very different temperature profile across the disk surface. 
The observed increase in size resembles that found from interband lag measurements when the BLR emission is not considered.

The idea that extremely large accretion disks are required to explain the microlensing durations was challenged by \cite{Fian2023}, who were able to show that the very large measured sizes are likely the result of strong DC emission from the inner parts of the BLR, regions that are far greater than the 2500\AA\ emitting part of the inner accretion disk. 
Here, I use the present model to calculate a 2500\AA\ SB map of Mrk\,817, which includes three components similar to the ones calculated for the L$_{\rm min}$ case: A central accretion disk, a BLR, and a disk wind. 
 The wind is somewhat different from the one considered earlier. It is observed over the range of 200-1500\Rg, it is made of numerous high column density clouds with 
$N_H \ge 10^{23.5}$\,cm$^{-2}$, and its covering factor is given by 
$dC \propto R^{-1} dR$. The total covering factor is 0.2, which makes the total wind emission similar to the wind considered earlier.
The wind is launched perpendicular to the disk's surface, so it does not shield the BLR.
The wind spectrum is similar to the one shown in Fig.~\ref{wind_no_wind_BLRs}, and the BLR spectrum is identical to the one shown in that diagram.

\begin{figure} \centering      
\includegraphics[width=0.95\linewidth]{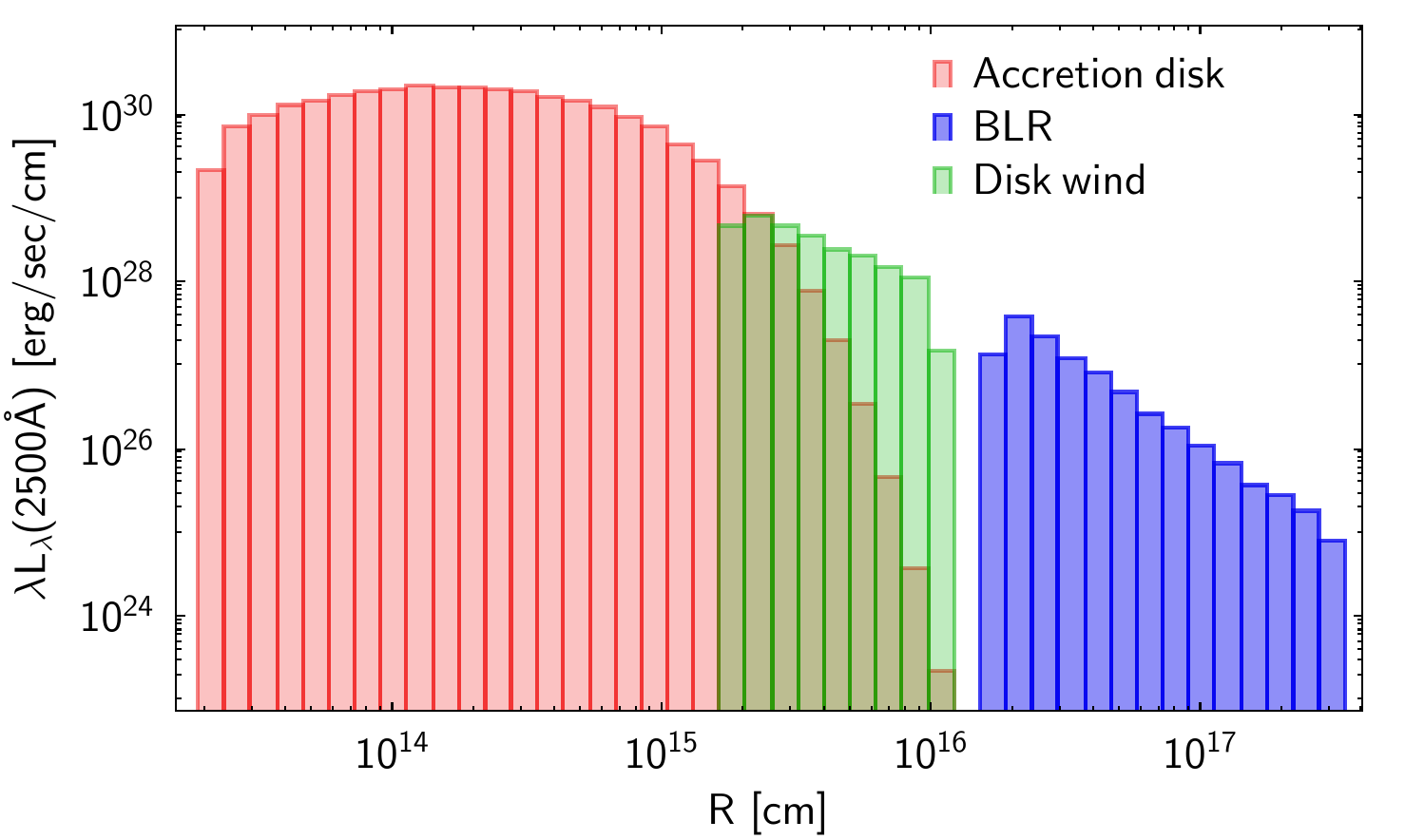}
        \caption{Surface brightness at 2500\AA\ for a there component model, which includes a central accretion disk, disk wing, and a BLR, with the properties described in the text. The listed luminosity is for equal-width rings over the entire range of distances. 
        }
        
        \label{disk_wind_BLR_2500}
\end{figure}

The SB of the three-component model is shown in Fig.~\ref{disk_wind_BLR_2500}. The total $\lambda L_{\lambda}$(2500\AA) 
is $9.8 \times 10^{43}$~\ergs, the relative luminosities are:
\begin{center}
L(disk):L(wind):L(BLR)=1:0.21:0.12, 
\end{center}
and the mean SB radii are:
R(disk)=110\Rg, R(wind)=760\Rg, and R(BLR)=8170\Rg. The mean SB radius is 9.4 times larger than R(disk). These calculations can be scaled up in BH mass and luminosity and are entirely consistent with 
 \cite{Fian2023} calculations. Thus, the global SB model of Mrk\,817 supports the suggestion that its disk size and temperature distribution are consistent with standard thin accretion disks.

\subsection{The variability timescales of AGN}

The variability timescales of AGN disks have been discussed in numerous publications (see section~2). Much of the older discussion is based on the suggestion that a DRW process (or, equivalently, a CAR(1) process, e.g., \cite{Kelly2009}) can describe the optical-UV variations. Here, I focus on two recent publications, 
\cite{Burke2021} and \cite{Arevalo2024}, which present a robust statistical analysis of high-cadence, long-duration light curves.

\cite{Burke2021} analyzed the variability of 67 AGN, out of a much larger sample, with BH mass in the range $10^4-10^{10}$\Msun. For most objects in the sample \MBH$>10^{6.5}$\Msun.
Based on earlier studies, they assumed that a DRW process could describe the PSD of the optical-UV variations with a characteristic time \tdamp\, which is determined from the break frequency in the power spectrum (or power spectral density, PSD) of the source.  Their analysis suggests a significant correlation between this time and the BH mass given by:
\begin{center}
    \tdamp=107M$_8 ^{0.38}$\,days, where M$_8$=\MBH/$10^8$\Msun.
\end{center}
According to the paper, the uncertainties in this expression are small enough to invert it and obtain a new way to measure \MBH\ with an intrinsic scatter similar to that of RM-based estimates. The common wavelength used in the analysis was 2500\AA, and the scaling of the damping time to this wavelength, based on \cite{MacLeod2010}, was
\tdamp$\propto \lambda_{\rm rest}^{0.17}$. The disk mean emissivity distance from the BH is R2500.

The \MBH-vs-R2500 correlation in \cite{Burke2021} is based on two ways to estimate R2500: from the part of the central accretion disk emitting this radiation and from microlensing measurements. According to the paper, the scatter in the measurements was too large to exclude one possibility or the other completely. As shown in the previous section and \cite{Fian2023}, the microlensing size measurements are strongly influenced by diffuse emission from disk winds, if they exist, and the extended BLR. Therefore, the discussion considers only 2500\AA\ emission from the central accretion disk.

The \cite{Arevalo2024} study is based on g-band photometry in a much larger sample with longer-duration light curves. The limited redshift range of 0.6-0.7 made it easier to choose a common wavelength for the analysis, 2900\AA, without additional assumptions about the dependence of the break frequency (or break time, $t_{\rm break}$) on wavelength. An important result of this study is the dependence of $t_{\rm break}$ on both BH mass and L/Ledd:
\begin{center}   
$ t_{\rm break} \propto $M$_8^{0.55-0.65}$(L/\Ledd)$^{0.3-0.35}$ \,\, days.
\end{center}
Another finding is that their best-fitting models have slopes above the power-spectrum break between 2.5 and 3, compared with the $f^{-2}$ dependence in DRW models. For the case of $f^{-2.5}$, M$_8$=1 and L/\Ledd=0.1, $t_{\rm break}$(2900\AA) is approximately 192 days. As shown below, this value of $t_{\rm break}$ agrees with the analysis of the Mrk\,817 observations.

The dynamical and thermal time scales for a standard optically thick geometrically thin accretion disk can be written as:
\begin{equation}
    t_{\rm dyn}=0.005 M_{8} \left [ \frac{R}{R_g} \right [ ^{3/2} \,\, {\rm days}
\end{equation}
and 
\begin{equation}
    t_{\rm th}=t_{\rm dyn}/\alpha \,\, ,
    \label{t_thermal}
\end{equation}
 where $\alpha$ is the viscosity parameter of the disk\footnote{ (\cite{Burke2021} used slightly different constants and the orbital time scale,
$t_{\rm orb}=2 \pi t_{\rm dyn}$) }.
Both times are very short compared to the viscous time of the disk $t_{\rm vis}$. 
The physical origin of the damping time scale could be associated with the thermal time scale at the radius where variability is driven. This agrees with earlier works that link the optical-UV damping time to the thermal time scale at the UV-emitting parts of the accretion disk. Unfortunately, this is hard to confirm in individual sources because \tdamp\ also depends on the BH spin through the accretion efficiency $\eta$  and the unknown viscosity parameter $\alpha$. 

Mrk\,817 is a unique laboratory for comparing observed and predicted variability timescales at several different wavelengths.
The source is one of the objects included in the \cite{Burke2021} sample with  \tdamp\ of approximately 30 days.
A PSD analysis based on the AGN STORM 2 \hst\ and \swift\ data (not shown here) suggests \tdamp(V-band)$\approx 30$ days, and \tdamp(1200-1700\AA)$\approx 16$ days, or $t_{\rm break}$ in the range 100-200 days. Assuming a standard SS73 disk, with the mass of the BH in Mrk\,817, and the accretion rate during the wind phase, I find that 50\% of the 1500\AA\ flux (representing the 1200-1700\AA\ range) is emitted between 24\Rg\ and 80\Rg. The 50\%\ flux range for the V-band is
110-430\Rg. This range of radii overlaps with the radii where most wind outflow occurs (section 2.3.1).

For magnetohydrodynamic simulation of AGN accretion disks,
$\alpha$ is typically 0.01, and for radiation-pressure dominated accretion disks, $\alpha$ is 5-10 times larger. 
Using  eqn.~\ref{t_thermal}, the thermal times associated with the 1500\AA\ radiations range from 12-60 days. The corresponding numbers associated with the V-band are in the range of 140 to 700 days.
Assuming that $t_{\rm break}$ is a reliable measure of the variability timescale, the observations at both wavelengths are consistent with the assumption that the variability timescale is the disk's local thermal time.

Given all this and the significant uncertainties associated with $t_{\rm th}$, I speculate that for Mrk\,817, there is a link between the location of the disk wind, which was determined from the need to explain the SED shape  (section~\ref{disk_winds}), the local thermal time, and the optical-UV variability timescales. A possible scenario is fluctuations driven by thermal instabilities associated with the radial dependence of the disk atmosphere's opacity.
Such fluctuations can induce line or continuum radiation pressure-driven mass loss from the disk's surface due to the local flux or the radiation from the inner regions of the disk. Thermal instabilities can also result in a reduced mass loss rate in other parts of the disks. Such fluctuations, whose properties require numerical simulations that are not yet available, can lead to color-dependent luminosity variations. 

\subsection{How realistic are disk-wind models?}

The original SN12 disk-wind model was constructed to explain the short-wavelength steepening of the spectrum of many AGN. The model assumed steady-state winds with a time-independent mass outflow rate.
A different model, by LD14, aimed to answer the same question. It suggests that the parts of the disk with temperatures close to 50,000K behave like the atmospheres of O-type stars with similar temperatures and $g_{\rm eff}$. Such stars produce fast stellar winds driven by line radiation pressure. The model proposed winds with similar properties over the hot parts of the disk, where the surface temperature is close to this value, and used observations of O-type stars to calculate the emitted flux over this range of temperatures. For the lower-mass BHs discussed in LD14, the steepening begins well outside the ISCO. This results in SED, which is practically independent of the black hole's spin. Like SN12, the paper discusses constant mass outflow rate winds.

An additional point discussed by the two papers is the energy associated with the wind's acceleration. LD14 distinguishes between two cases. One where all the energy is taken from the local radiation field, and the wind reaches the local escape velocity, and one that is independent of the local flux. The second scenario results in a failed wind and gas that returns to the disk, but not necessarily along the original flow lines. SN12 assumed that the absorbed energy is half the radiation needed to reach the escape velocity. This resulted in a failed wind too. The effect of all this on the observed flux depends, among other things, on the observer's line of sight to the source. Perhaps more important is the fact that the LD14 disk wind is unavoidable for disks around BH with mass smaller than about $10^{9}$\Msun\ and L/\Ledd$>0.01$, which means that for such objects, one never observes a disk SED similar to the one suggested in SS73.

The case discussed in this paper is different. It assumes a short, repetitive wind with a local mass outflow rate that changes on a timescale of about 100 days. The steepest (i.e., softest) SED is observed during the highest mass outflow rate. Additionally, the sources of acceleration and the wind's kinetic energy were not addressed. However, the example discussed in section~2.4 refers to wind velocity that exceeds the escape velocity at its observed location. 

A better understanding of the AGN disk-wind model requires more sophisticated numerical calculations and additional observational verifications.  On top of the observational signatures discussed in this paper (SED shape, variability time scale, the spectrum of the photoionized wind), one can examine the equivalent width of high ionization lines like \civ\ and \heii, and line ratios like \heii/\La, that can be used to estimate the ionizing luminosity of the source and compare to model predictions. Several attempts have been made to address these issues in the past \citep[e.g.,][]{Bonning2013}. Still, their calculated broad-line luminosities differed significantly from those presented in this work, and their narrow-line ratios did not account for dust in the narrow line region. One can also map, in more detail, the wavelength-dependent variability timescale in Mrk\,817, and other AGN.

\subsection{Implications for other AGN}

Most of the analysis presented in this paper applies to Mrk\,817, and the superb data obtained during the AGN\,STORM\,2 campaign. This source is not unusual in terms of its spectral properties, BH mass, line and continuum variations, and the size of the BLR. It is thus important to search for similar observational indications for disk winds in other sources to verify whether the above conclusions apply to the entire AGN population.

The best indicators of disk-wind systems are the relationships between black hole mass, variability timescale, and mass outflow rate. The first two are already known for a large number of AGN. Examples are the references mentioned above, particularly the data presented in \cite{Arevalo2024}, where many sources, with a large range in BH mass and L/\Ledd,  and long-duration light curves, are available. More detailed spectral analysis of such objects would be most useful, especially in cases where the mass outflow rate can be measured from X-ray observations over hundreds of days. The fundamental requirement is a changing UV-optical SED shape during bolometric luminosity variations. This can be looked for using \hst\ observations in low redshift AGN, and ground-based spectroscopy in high redshift AGN. Fitting combined optical-UV spectra with accretion disk spectra can be used to verify the models. There are already a large number of high-z high-L AGN where this has been attempted \citep[see e.g.,][]{Capellupo2015, Capellupo2016} but confirmation from higher or lower luminosity epochs is still missing.
 High-resolution X-ray spectroscopy, of sufficiently high quality to measure wind velocity, will be most helpful in this respect.

  Measuring the variability timescale in high-luminosity, large BH mass AGN is challenging since the thermal time increases with $M^{1/2}$.  A careful line profile analysis is another way to search for outflow signatures in such sources.
  Studies of hundreds of high-L AGN \citep[see][and references therein]{Matthews2023,Temple2021} resulted in the claim that disk outflows are very common in such objects. The winds are detected much further from the central BH, at distances similar to the distance of the BLR. This idea can be verified by detailed 2D-RM of strong emission lines, where the different kinematics of the lines' core (rotation) and wings (outflow) reflect different velocity fields.

\section{Conclusions}

This paper addresses accretion disk winds and the associated spectral variations in AGN. 
A detailed disk-wind-BLR model is presented and compared to recent observations of Mrk\,817. The main results can be summarized as follows:
\begin{itemize}
\item 
A powerful wind ejected from the surface of the accretion disk in Mrk\,817 changes the mass accretion rate onto the BH. A wind with a mean mass outflow rate of 0.075\Msun/yr, with a half ejection radius of 50\Rg\ and a characteristic variability timescale of about 100 days, can explain the observed changes in luminosity and SED shape of the source. The bolometric luminosity change due to the wind was about a factor of three, and the corresponding reduction in ionizing luminosity was approximately a factor of four.

\item 
The wind observed in Mrk\,817 is probably ejected perpendicular to the disk surface and extends over a large range of radii, from about 50 to 1000\Rg, without shielding the BLR.

\item 
The diffuse emission from the BLR during the highest and lowest luminosity phases, combined with emission by the wind material, can explain the observed spectral variations in this source, the time lags of the strong emission lines, and the observed interband lags. There are additional implications for microlensing size measurements in more luminous AGN.

\item
The variability timescales of the 1500-5500\AA\ radiation, the thermal time of the disk at the locations of peak emission of these wavelengths, and the range of radii over which most of the mass outflow is taking place are all in agreement with the observed BH mass and accretion rate of Mrk\,817. This suggests a general connection between disk winds, SED shape, thermal timescales, and optical-UV variability in AGN.

\end{itemize}

\section*{Acknowledgements}

I thank Ari Laor and Shai Kaspi for their useful comments and suggestions, and an anonymous referee for suggesting that the discussion about implications for other AGN be expanded.
This work would not have been possible without the extraordinary data collected by the AGN STORM 2 collaboration. I am deeply grateful to the initiators, leaders, and all other contributors for allowing me to participate in this unique research project.

\section*{Data Availability}
All data used in this paper were obtained from published  AGN STORM 2 papers
\citep[see][]{Kara2021,Cackett2023,Homayouni2023,Partington2023,Homayouni2024,Lewin2024,Zaidouni2024,Netzer2024} where links to \hst\ and \swift\ data are provided.


\begin{thebibliography}{}
\makeatletter
\relax
\def\mn@urlcharsother{\let\do\@makeother \do\$\do\&\do\#\do\^\do\_\do\%\do\~}
\def\mn@doi{\begingroup\mn@urlcharsother \@ifnextchar [ {\mn@doi@}
  {\mn@doi@[]}}
\def\mn@doi@[#1]#2{\def\@tempa{#1}\ifx\@tempa\@empty \href
  {http://dx.doi.org/#2} {doi:#2}\else \href {http://dx.doi.org/#2} {#1}\fi
  \endgroup}
\def\mn@eprint#1#2{\mn@eprint@#1:#2::\@nil}
\def\mn@eprint@arXiv#1{\href {http://arxiv.org/abs/#1} {{\tt arXiv:#1}}}
\def\mn@eprint@dblp#1{\href {http://dblp.uni-trier.de/rec/bibtex/#1.xml}
  {dblp:#1}}
\def\mn@eprint@#1:#2:#3:#4\@nil{\def\@tempa {#1}\def\@tempb {#2}\def\@tempc
  {#3}\ifx \@tempc \@empty \let \@tempc \@tempb \let \@tempb \@tempa \fi \ifx
  \@tempb \@empty \def\@tempb {arXiv}\fi \@ifundefined
  {mn@eprint@\@tempb}{\@tempb:\@tempc}{\expandafter \expandafter \csname
  mn@eprint@\@tempb\endcsname \expandafter{\@tempc}}}

\bibitem[\protect\citeauthoryear{{Ar{\'e}valo}, {Churazov}, {Lira},
  {S{\'a}nchez-S{\'a}ez}, {Bernal}, {Hern{\'a}ndez-Garc{\'\i}a},
  {L{\'o}pez-Navas}  \& {Patel}}{{Ar{\'e}valo} et~al.}{2024}]{Arevalo2024}
{Ar{\'e}valo} P.,  {Churazov} E.,  {Lira} P.,  {S{\'a}nchez-S{\'a}ez} P.,
  {Bernal} S.,  {Hern{\'a}ndez-Garc{\'\i}a} L.,  {L{\'o}pez-Navas} E.,
  {Patel} P.,  2024, \mn@doi [\aap] {10.1051/0004-6361/202347080}, \href
  {https://ui.adsabs.harvard.edu/abs/2024A&A...684A.133A} {684, A133}

\bibitem[\protect\citeauthoryear{{Bonning}, {Shields}, {Stevens}  \&
  {Salviander}}{{Bonning} et~al.}{2013}]{Bonning2013}
{Bonning} E.~W.,  {Shields} G.~A.,  {Stevens} A.~C.,   {Salviander} S.,  2013,
  \mn@doi [\apj] {10.1088/0004-637X/770/1/30}, \href
  {https://ui.adsabs.harvard.edu/abs/2013ApJ...770...30B} {770, 30}

\bibitem[\protect\citeauthoryear{{Burke} et~al.,}{{Burke}
  et~al.}{2021}]{Burke2021}
{Burke} C.~J.,  et~al., 2021, \mn@doi [Science] {10.1126/science.abg9933},
  \href {https://ui.adsabs.harvard.edu/abs/2021Sci...373..789B} {373, 789}

\bibitem[\protect\citeauthoryear{{Cackett} et~al.,}{{Cackett}
  et~al.}{2023}]{Cackett2023}
{Cackett} E.~M.,  et~al., 2023, \mn@doi [\apj] {10.3847/1538-4357/acfdac},
  \href {https://ui.adsabs.harvard.edu/abs/2023ApJ...958..195C} {958, 195}

\bibitem[\protect\citeauthoryear{{Capellupo}, {Netzer}, {Lira}, {Trakhtenbrot}
  \& {Mej{\'{\i}}a-Restrepo}}{{Capellupo} et~al.}{2015}]{Capellupo2015}
{Capellupo} D.~M.,  {Netzer} H.,  {Lira} P.,  {Trakhtenbrot} B.,
  {Mej{\'{\i}}a-Restrepo} J.,  2015, \mn@doi [\mnras] {10.1093/mnras/stu2266},
  \href {http://adsabs.harvard.edu/abs/2015MNRAS.446.3427C} {446, 3427}

\bibitem[\protect\citeauthoryear{{Capellupo}, {Netzer}, {Lira}, {Trakhtenbrot}
  \& {Mej{\'{\i}}a-Restrepo}}{{Capellupo} et~al.}{2016}]{Capellupo2016}
{Capellupo} D.~M.,  {Netzer} H.,  {Lira} P.,  {Trakhtenbrot} B.,
  {Mej{\'{\i}}a-Restrepo} J.,  2016, \mn@doi [\mnras] {10.1093/mnras/stw937},
  \href {http://adsabs.harvard.edu/abs/2016MNRAS.460..212C} {460, 212}

\bibitem[\protect\citeauthoryear{{Cornachione} \& {Morgan}}{{Cornachione} \&
  {Morgan}}{2020}]{Cornachione2020}
{Cornachione} M.~A.,  {Morgan} C.~W.,  2020, \mn@doi [\apj]
  {10.3847/1538-4357/ab8aed}, \href
  {https://ui.adsabs.harvard.edu/abs/2020ApJ...895...93C} {895, 93}

\bibitem[\protect\citeauthoryear{{De Rosa} et~al.,}{{De Rosa}
  et~al.}{2015}]{deRosa2015}
{De Rosa} G.,  et~al., 2015, \mn@doi [\apj] {10.1088/0004-637X/806/1/128},
  \href {https://ui.adsabs.harvard.edu/abs/2015ApJ...806..128D} {806, 128}

\bibitem[\protect\citeauthoryear{{Ferland} et~al.,}{{Ferland}
  et~al.}{2017}]{Ferland2017}
{Ferland} G.~J.,  et~al., 2017, \rmxaa, \href
  {https://ui.adsabs.harvard.edu/abs/2017RMxAA..53..385F} {53, 385}

\bibitem[\protect\citeauthoryear{Fian, Chelouche  \& Kaspi}{Fian
  et~al.}{2023}]{Fian2023}
Fian C.,  Chelouche D.,   Kaspi S.,  2023, \mn@doi [Astronomy &amp;
  Astrophysics] {10.1051/0004-6361/202346766}, 677, A94

\bibitem[\protect\citeauthoryear{{Homayouni} et~al.,}{{Homayouni}
  et~al.}{2023}]{Homayouni2023}
{Homayouni} Y.,  et~al., 2023, \mn@doi [\apj] {10.3847/1538-4357/acc45a}, \href
  {https://ui.adsabs.harvard.edu/abs/2023ApJ...948...85H} {948, 85}

\bibitem[\protect\citeauthoryear{{Homayouni} et~al.,}{{Homayouni}
  et~al.}{2024}]{Homayouni2024}
{Homayouni} Y.,  et~al., 2024, \mn@doi [\apj] {10.3847/1538-4357/ad1be4}, \href
  {https://ui.adsabs.harvard.edu/abs/2024ApJ...963..123H} {963, 123}

\bibitem[\protect\citeauthoryear{{Kara} et~al.,}{{Kara}
  et~al.}{2021}]{Kara2021}
{Kara} E.,  et~al., 2021, \mn@doi [\apj] {10.3847/1538-4357/ac2159}, \href
  {https://ui.adsabs.harvard.edu/abs/2021ApJ...922..151K} {922, 151}

\bibitem[\protect\citeauthoryear{Kelly, Bechtold  \& Siemiginowska}{Kelly
  et~al.}{2009}]{Kelly2009}
Kelly B.~C.,  Bechtold J.,   Siemiginowska A.,  2009, \mn@doi [The
  Astrophysical Journal] {10.1088/0004-637X/698/1/895}, 698, 895

\bibitem[\protect\citeauthoryear{{Korista} et~al.,}{{Korista}
  et~al.}{1995}]{Korista1995}
{Korista} K.~T.,  et~al., 1995, \mn@doi [\apjs] {10.1086/192144}, \href
  {https://ui.adsabs.harvard.edu/abs/1995ApJS...97..285K} {97, 285}

\bibitem[\protect\citeauthoryear{{Laor} \& {Davis}}{{Laor} \&
  {Davis}}{2014}]{Laor2014}
{Laor} A.,  {Davis} S.~W.,  2014, \mn@doi [\mnras] {10.1093/mnras/stt2408},
  \href {http://adsabs.harvard.edu/abs/2014MNRAS.438.3024L} {438, 3024}

\bibitem[\protect\citeauthoryear{{Lewin} et~al.,}{{Lewin}
  et~al.}{2024}]{Lewin2024}
{Lewin} C.,  et~al., 2024, \mn@doi [\apj] {10.3847/1538-4357/ad6b08}, \href
  {https://ui.adsabs.harvard.edu/abs/2024ApJ...974..271L} {974, 271}

\bibitem[\protect\citeauthoryear{{MacLeod} et~al.,}{{MacLeod}
  et~al.}{2010}]{MacLeod2010}
{MacLeod} C.~L.,  et~al., 2010, \mn@doi [\apj] {10.1088/0004-637X/721/2/1014},
  \href {https://ui.adsabs.harvard.edu/abs/2010ApJ...721.1014M} {721, 1014}

\bibitem[\protect\citeauthoryear{{Maoz} et~al.,}{{Maoz}
  et~al.}{1993}]{Maoz1993}
{Maoz} D.,  et~al., 1993, \mn@doi [\apj] {10.1086/172310}, \href
  {https://ui.adsabs.harvard.edu/abs/1993ApJ...404..576M} {404, 576}

\bibitem[\protect\citeauthoryear{{Matthews} et~al.,}{{Matthews}
  et~al.}{2023}]{Matthews2023}
{Matthews} J.~H.,  et~al., 2023, \mn@doi [\mnras] {10.1093/mnras/stad2895},
  \href {https://ui.adsabs.harvard.edu/abs/2023MNRAS.526.3967M} {526, 3967}

\bibitem[\protect\citeauthoryear{{Morgan}, {Hyer}, {Bonvin}, {Mosquera},
  {Cornachione}, {Courbin}, {Kochanek}  \& {Falco}}{{Morgan}
  et~al.}{2018}]{Morgan2018}
{Morgan} C.~W.,  {Hyer} G.~E.,  {Bonvin} V.,  {Mosquera} A.~M.,  {Cornachione}
  M.,  {Courbin} F.,  {Kochanek} C.~S.,   {Falco} E.~E.,  2018, \mn@doi [\apj]
  {10.3847/1538-4357/aaed3e}, \href
  {https://ui.adsabs.harvard.edu/abs/2018ApJ...869..106M} {869, 106}

\bibitem[\protect\citeauthoryear{{Motta}, {Mediavilla}, {Rojas}, {Falco},
  {Jim{\'e}nez-Vicente}  \& {Mu{\~n}oz}}{{Motta} et~al.}{2017}]{Motta2017}
{Motta} V.,  {Mediavilla} E.,  {Rojas} K.,  {Falco} E.~E.,
  {Jim{\'e}nez-Vicente} J.,   {Mu{\~n}oz} J.~A.,  2017, \mn@doi [\apj]
  {10.3847/1538-4357/835/2/132}, \href
  {https://ui.adsabs.harvard.edu/abs/2017ApJ...835..132M} {835, 132}

\bibitem[\protect\citeauthoryear{{Netzer}}{{Netzer}}{2022}]{Netzer2022}
{Netzer} H.,  2022, \mn@doi [\mnras] {10.1093/mnras/stab3133}, \href
  {https://ui.adsabs.harvard.edu/abs/2022MNRAS.509.2637N} {509, 2637}

\bibitem[\protect\citeauthoryear{{Netzer} et~al.,}{{Netzer}
  et~al.}{2024}]{Netzer2024}
{Netzer} H.,  et~al., 2024, \mn@doi [\apj] {10.3847/1538-4357/ad8160}, \href
  {https://ui.adsabs.harvard.edu/abs/2024ApJ...976...59N} {976, 59}

\bibitem[\protect\citeauthoryear{{Partington} et~al.,}{{Partington}
  et~al.}{2023}]{Partington2023}
{Partington} E.~R.,  et~al., 2023, \mn@doi [\apj] {10.3847/1538-4357/acbf44},
  \href {https://ui.adsabs.harvard.edu/abs/2023ApJ...947....2P} {947, 2}

\bibitem[\protect\citeauthoryear{{Scott}, {Kriss}, {Brotherton}, {Green},
  {Hutchings}, {Shull}  \& {Zheng}}{{Scott} et~al.}{2004}]{Scott2004}
{Scott} J.~E.,  {Kriss} G.~A.,  {Brotherton} M.,  {Green} R.~F.,  {Hutchings}
  J.,  {Shull} J.~M.,   {Zheng} W.,  2004, \mn@doi [\apj] {10.1086/422336},
  \href {https://ui.adsabs.harvard.edu/abs/2004ApJ...615..135S} {615, 135}

\bibitem[\protect\citeauthoryear{{Shakura} \& {Sunyaev}}{{Shakura} \&
  {Sunyaev}}{1973}]{Shakura1973}
{Shakura} N.~I.,  {Sunyaev} R.~A.,  1973, \aap, \href
  {http://adsabs.harvard.edu/abs/1973A%26A....24..337S} {24, 337}

\bibitem[\protect\citeauthoryear{{Slone} \& {Netzer}}{{Slone} \&
  {Netzer}}{2012}]{Slone2012}
{Slone} O.,  {Netzer} H.,  2012, \mn@doi [\mnras]
  {10.1111/j.1365-2966.2012.21699.x}, \href
  {http://adsabs.harvard.edu/abs/2012MNRAS.426..656S} {426, 656}

\bibitem[\protect\citeauthoryear{{Stevans}, {Shull}, {Danforth}  \&
  {Tilton}}{{Stevans} et~al.}{2014}]{Stevans2014}
{Stevans} M.~L.,  {Shull} J.~M.,  {Danforth} C.~W.,   {Tilton} E.~M.,  2014,
  \mn@doi [\apj] {10.1088/0004-637X/794/1/75}, \href
  {https://ui.adsabs.harvard.edu/abs/2014ApJ...794...75S} {794, 75}

\bibitem[\protect\citeauthoryear{Sun, Wang, Chen  \& Zheng}{Sun
  et~al.}{2014}]{Sun2014}
Sun Y.-H.,  Wang J.-X.,  Chen X.-Y.,   Zheng Z.-Y.,  2014, \mn@doi [The
  Astrophysical Journal] {10.1088/0004-637X/792/1/54}, 792, 54

\bibitem[\protect\citeauthoryear{Temple, Ferland, Rankine, Chatzikos  \&
  Hewett}{Temple et~al.}{2021}]{Temple2021}
Temple M.~J.,  Ferland G.~J.,  Rankine A.~L.,  Chatzikos M.,   Hewett P.~C.,
  2021, \mn@doi [Monthly Notices of the Royal Astronomical Society]
  {10.1093/mnras/stab1610}, 505, 3247

\bibitem[\protect\citeauthoryear{{Vanden Berk} et~al.,}{{Vanden Berk}
  et~al.}{2004}]{Vandenberk2004}
{Vanden Berk} D.~E.,  et~al., 2004, \mn@doi [\apj] {10.1086/380563}, \href
  {https://ui.adsabs.harvard.edu/abs/2004ApJ...601..692V} {601, 692}

\bibitem[\protect\citeauthoryear{{Wamsteker} et~al.,}{{Wamsteker}
  et~al.}{1990}]{Wamsteker1990}
{Wamsteker} W.,  et~al., 1990, \mn@doi [\apj] {10.1086/168707}, \href
  {https://ui.adsabs.harvard.edu/abs/1990ApJ...354..446W} {354, 446}

\bibitem[\protect\citeauthoryear{{Zaidouni} et~al.,}{{Zaidouni}
  et~al.}{2024}]{Zaidouni2024}
{Zaidouni} F.,  et~al., 2024, \mn@doi [\apj] {10.3847/1538-4357/ad6771}, \href
  {https://ui.adsabs.harvard.edu/abs/2024ApJ...974...91Z} {974, 91}

\makeatother
\end{thebibliography}


\bsp	
\label{lastpage}
\end{document}